\documentclass[10pt,letterpaper,twocolumn]{article} 

\usepackage{ol2}
\usepackage[draft]{hyperref}
\usepackage{amsmath}
\begin{document}

\twocolumn[ 

\title{Continuous-wave non-classical light with GHz squeezing bandwidth}


\author{Stefan Ast$^1$, Aiko Samblowski$^1$, Moritz Mehmet$^{1,2}$, Sebastian Steinlechner$^1$, Tobias Eberle$^{1,2}$ and Roman Schnabel$^{1,*}$}

\address{
$^1$Max-Planck-Institute for Gravitational Physics, Albert-Einstein-Institute, and Institut f\"{u}r Gravitationsphysik,
Leibniz Universit\"{a}t Hannover, Callinstrasse 38, D-30167 Hannover, Germany
\\
$^2$Centre for Quantum Engineering and Space-Time Research - QUEST, Leibniz Universit\"{a}t Hannover, Welfengarten 1, 30167 Hannover, Germany
\\
$^*$Corresponding author: roman.schnabel@aei.mpg.de
}

\begin{abstract}
Squeezed states can be employed for entanglement-based continuous-variable quantum key distribution, where the secure key 
rate is proportional to the bandwidth of the squeezing. We produced a non-classical continuous-wave laser field at the  
telecommunication wavelength of 1550\,nm, which showed squeezing over a bandwidth of more than 2\,GHz. 
The experimental setup used parametric down-conversion via a periodically poled potassium titanyl phosphate crystal (PPKTP). 
We did not use any resonant enhancement for the fundamental wavelength, which should in principle allow a production of squeezed light
 over the full phase-matching bandwidth of several nanometers. We measured the squeezing to be up to 0.3\,dB below the vacuum noise 
from 50\,MHz to 2\,GHz limited by the measuring bandwidth of the homodyne detector. The squeezing strength 
was possibly limited by thermal lensing inside the non-linear crystal. 
\end{abstract}

\ocis{270.6570}]

\begin{figure}[tbp]
\centerline{\includegraphics[width=8.6cm]{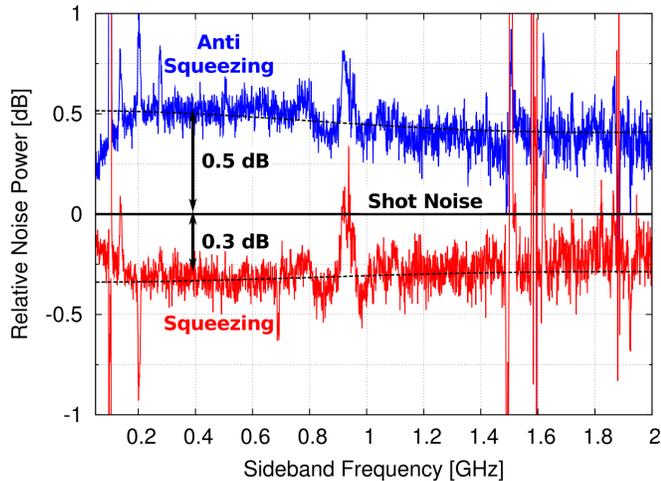}}
\caption{Quantum noise spectra normalized to shot noise.  
 The traces shown include a dark noise correction, which allows a comparison with numerical simulations of our system 
(dashed lines) and which marginally improves the squeezing data only above 1\,GHz. A non-classical noise suppression 
(squeezing) of up to 0.3\,dB and 0.5\,dB of anti-squeezing are observed from 0.05\,-\,2\,GHz.
}
\label{fig:GHz-Sqz}
\end{figure}

\noindent
Squeezed states of light are non-classical states, capable of being the resource for entanglement generation in the continuous variable
(CV) regime \cite{Yuen78, Ou1992, PhysRevA.61.010303, Bowen2003,Diguglielmo2007}. Possible approaches are to either overlap 
two independent squeezed modes on a beam splitter, or mix  single-mode squeezing with vacuum. 
The quadratures of the beam splitter's outputs will then be entangled being an effective toolbox for quantum key distribution (QKD)
\cite{springerlink:10.1007/s11080-007-9030-x, 0295-5075-87-2-20005, QKD-Werner}.
The data rate for entanglement based QKD with squeezed states increases with the squeezing strength and is proportional to the squeezing
bandwidth. While the first increases the average number of bits per measurement, the latter will allow a
higher measuring speed. Thus both increase the key rate. One approach for applications of QKD is the distribution of  
continuous-wave CV entangled states via standard telecom fibers \cite{PhysRevA.76.042305,Momo2010}. 
Here, the wavelength of 1550\,nm is advantageous since it offers low optical loss, which is necessary to  protect the entanglement 
against decoherence \cite{PhysRevLett.102.130501}. In a km scale fiber network, however, decoherence will still be the limiting factor and the squeezing bandwidth is more robust than the squeezing strength.\\
The generation of squeezed states of light has first been demonstrated in 1985 by Slusher \emph{et al.} \cite{Slu85}. 
Today, a common method to produce continuous-wave squeezing is parametric down-conversion 
in second-order nonlinear crystals placed inside optical resonators \cite{PhysRevA.29.408,PhysRevLett.57.2520}. 
The resonator enhances the fundamental field to be squeezed, allowing non-classical noise suppressions of more than 10\,dB in 
recent experiments \cite{Vahlbruch2008,PhysRevA.81.013814,Eberle2010}, but also limits its bandwidth. 
Without an optical resonator this limitation is not present, which is a typical situation for squeezed states in the pulsed laser regime \cite{SBSSL98}.
The highest bandwidth 
for continuous-wave squeezed light so far was measured in \cite{PhysRevA.81.013814} where 100\,MHz and a non-classical noise suppression of up to 11.5\,dB was achieved.\\
In the work presented here, no resonant enhancement of the fundamental wavelength was used. The squeezing bandwidth was thus 
limited by the phase-matching condition of the non-linear crystal. The consequences are a very high pump threshold 
for parametric oscillation and small squeezing factors achievable. We
partially compensated the absence of the resonator at the fundamental wavelength by resonantly enhancing the pump light instead. 
As the result we present the first measurement of a continuous-wave squeezed vacuum state with 2\,GHz bandwidth. 
The non-classical noise suppression was up to 0.3\,dB below shot noise.
Note that in the regime of squeezed laser \emph{pulses} \cite{PhysRevLett.59.2566,Dong:08} enhancement resonators are generally not necessary and also GHz squeezing bandwidths are possible, however, not demonstrated so far to the best of our knowledge.\\
\begin{figure}[!htbp]
\centerline{\includegraphics[width=8.6cm]{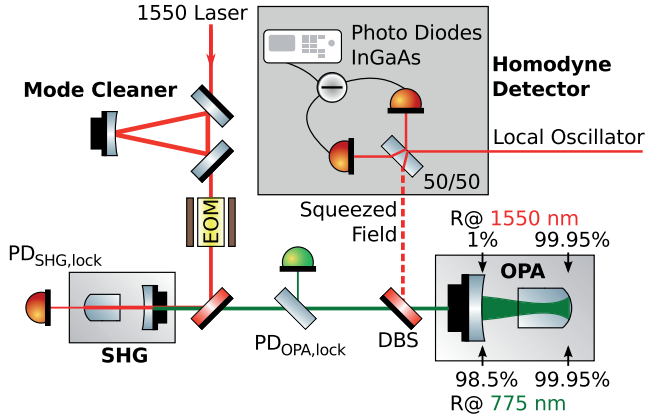}}
\caption{Schematic of the experimental setup. Mode Cleaner: three mirror spatial filter cavity; SHG: second harmonic generation cavity
 with nonlinear \mbox{PPKTP} crystal; OPA: squeezing resonator with nonlinear \mbox{PPKTP} crystal and high squeezing bandwidth;
DBS: dichroic beam splitter to separate 775\,nm and 1550\,nm light; $\text{PD}_{\text{SHG/OPA,lock}}$: photo detector
 measuring error signals for cavity length stabilization with 138\,MHz demodulation frequency; EOM: electro-optic modulator producing
 sidebands of 138\,MHz for the cavity length stabilization.}
\label{fig:GHz-Sqz-1550-exp-setup}
\end{figure}
In the experimental setup (Fig. \ref{fig:GHz-Sqz-1550-exp-setup}) 1550\,nm continuous-wave light from an erbium-doped fiber laser was used.
 First, the laser light was transmitted through a three mirror ring resonator for spatial mode cleaning. 1.1\,W were mode-matched into a 
second harmonic generation (SHG) cavity containing a \mbox{PPKTP} crystal. We used the SHG cavity described in \cite{Ast-SHG},
which converted 95\,\% of the 1550\,nm laser light into 775\,nm pump light for the squeezing resonator.\\
The squeezed-light source used another PPKTP crystal having a dimension of $1\times2\times9.3\,\text{mm}^{3}$ and a quasi phase-matching temperature of about 
$46\,^{\circ}\text{C}$. Its one end surface had a radius of curvature of $\text{RoC}=12\,\text{mm}$ and a highly reflective coating ($\text{HR}=99.95\,\%$) 
for fundamental as well as for the harmonic wavelength. The crystal's other surface was plano
and anti-reflective coated (AR $< 0.1\,\%$) 
to minimize intra-cavity loss.
A curved mirror ($\text{RoC}=25\,\text{mm}$) with reflectivities $\text{R}_{1550}< 1\,\%$
and $\text{R}_{775}=98.5\,\%$ was placed
24\,mm in front of the crystal, leading to a free spectral range of 3.67\,\text{GHz}. 
The crystal's end surface and the external mirror formed the linear cavity with a Finesse of
about $\mathcal{F}_{\text{775}}=350$ for the harmonic pump field. 
We stabilized the squeezing resonator length with the pump field by using a Pound Drever
Hall (PDH) scheme. Here the phase modulation applied before the SHG was down-converted and used to generate an error signal.
Due to the high Finesse at the pump wavelength, an intra-cavity pump power of up to 30\,W was achieved limited by thermal lensing inside the crystal. 
The waist size without thermal lensing was  $\text{w}_{0,775}=27\,\mu\text{m}$.
The maximum pump power was still far below the threshold for optical parametric oscillation (of about 1\,kW), thus limiting the achievable squeezing.\\  
\begin{figure}[t]
 \includegraphics[width=8.6cm]{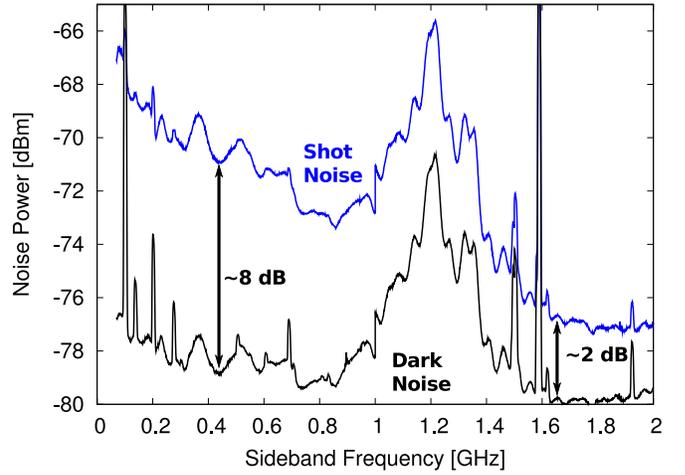}
\caption{Balanced homodyne detector characterization from 50\,MHz to 2\,GHz. Top: Homodyne detector shot noise at 0.75\,mW local oscillator power. 
Bottom: Homodyne detector dark noise. The dark noise distance is ranging from about 9\,dB at 50\,MHz, 
 8\,dB at 500\,MHz, and slowly degrading to about 2\,dB at 2\,GHz.}
\label{fig:GHz-Homo} 
\end{figure}
\begin{figure}[t]
\includegraphics[width=8.6cm]{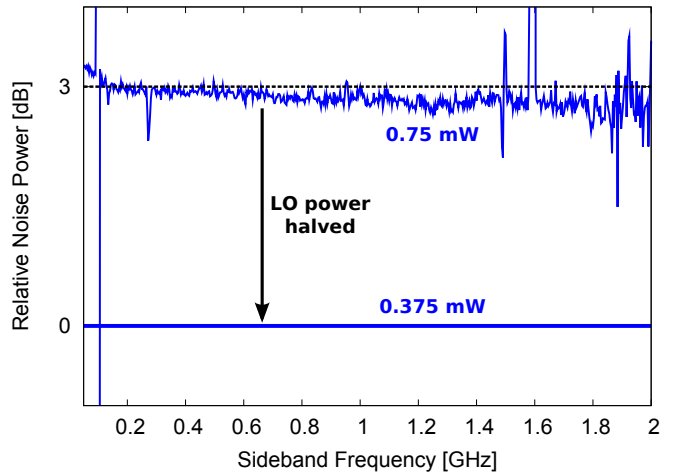}
\caption{Homodyne detector linearity. Measured shot noise
levels at 0.75\,mW (top) and 0.375\,mW (bottom) local oscillator
(LO) powers, both dark noise corrected and normalized to the data of the
latter. At these LO power levels the homodyne detector
is linear since halving the LO power results in a 3\,dB reduction
of the shot noise. A small saturation effect
is revealed towards higher frequencies. 
}
\label{fig:GHz-Homo-norm}
\end{figure} 
Our balanced homodyne detector (BHD) consisted of two FCI-H250G-InGaAs-70 (\emph{OSI Optoelectronics}) photo detectors with build-in transimpedance amplifiers, which
allowed us to measure up to 2\,GHz. A subtraction of the measured signals was achieved via commercially available power splitters within a 
frequency range of 50\,MHz to 1\,GHz and of 1\,GHz to 2\,GHz, respectively. The detector's dark noise clearance from the shot-noise level  at 
0.75\,mW local oscillator power was about 8\,dB over a frequency range of several hundred MHz, slowly degrading to 2\,dB towards 2\,GHz (Fig. \ref{fig:GHz-Homo}). 
The dark noise was measured with all BHD inputs blocked. The shot noise level was measured with only the signal input blocked. Several electronic pick-up peaks showed up in the homodyne detector's dark noise and the shot noise. There are modulation
 peaks from the PDH locking frequencies at 101\,MHz, 138\,MHz as well as their harmonics. The peaks between 1\,GHz and 2\,GHz
 can be seen in the detector dark noise and correspond to external electronic pick up noise. The broad peak structure around 1.2\,GHz is due to a resonance in the electronic circuit of the detector.
The homodyne detector spectrum was linear at a local oscillator (LO) power of up to 0.75\,mW (Fig. \ref{fig:GHz-Homo-norm}). Saturation effects
start to appear above 0.7\,GHz as the curve is slightly lower than 3\,dB. This effect degraded the measured squeezing strength, but only by negligible amounts. \\
We measured a squeezing level of up to 0.3\,dB and anti-squeezing of up to 0.5\,dB from 50\,MHz to 2\,GHz using an (intra-cavity) harmonic pump power of about 30\,W and a LO power of 0.75\,mW (Fig. \ref{fig:GHz-Sqz}). 
The peaks in the normalized spectrum also arise in the detector's dark noise and can be associated with electronic pick-up noise.
The squeezing strength is mainly limited by insufficient pump power far below the parametric oscillation threshold. 
When we further increased the input pump power, we did not observe a corresponding increase of the intra-cavity pump power. Neither did the cavity transmitted light power increase nor did the squeezing factor improve.
 From these observations we conclude that thermal lensing inside the crystal reduced the spatial overlap between the pump field and the fundamental field being mode-matched to the BHD.\\
In conclusion, we produced a continuous-wave squeezed vacuum field with a measured squeezing bandwidth of 2\,GHz. Our result serves
as a proof of principle experiment for the generation of a GHz bandwidth two-mode squeezed states for high-speed 
entanglement-based QKD. The bandwidth of our squeezing source was possibly only limited by the crystal's phase-matching 
bandwidth, however, the actually observed squeezing spectrum was considerably narrower and limited by the 2\,GHz detection 
bandwidth of the homodyne detector. The measured squeezing strength was just up to 0.3\,dB below shot noise, which is too 
low for efficient entanglement-based QKD. The conclusion from our experiment is thus to find a trade-off between the 
bandwidth and squeezing strength of the source. This trade-off can be realized by applying an enhancement resonator with a 
large linewidth of several GHz at the fundamental wavelength. First numerical simulations indicate that a low-finesse PPKTP 
standing-wave resonator of a few millimeter length should provide more than 3\,dB of squeezing over a 1\,GHz bandwidth.\\
We acknowledge support from the International Max Planck Research School on Gravitational Wave Astronomy and the EU FP 7 project Q-ESSENCE.
The authors like to thank Henning Kaufer, Oliver Gerberding and Vitus H\"{a}ndchen for useful discussions.

\end{document}